\documentclass[aps,prc,twocolumn,showpacs,superscriptaddress]{revtex4-1}
\usepackage{graphicx}
\usepackage{amsmath}
\usepackage{physics}
\usepackage{color}

\newcommand {\nc} {\newcommand}
\nc {\beq} {\begin{eqnarray}} \nc {\eol} {\nonumber \\} \nc {\eeq}
{\end{eqnarray}} \nc {\eeqn} [1] {\label{#1} \end{eqnarray}} \nc
{\eoln} [1] {\label{#1} \\} \nc {\ve} [1] {\mbox{\boldmath $#1$}}
\nc {\rref} [1] {(\ref{#1})} \nc {\Eq} [1] {Eq.~(\ref{#1})} \nc
{\re} [1] {Ref.~\cite{#1}} \nc {\dem} {\mbox{$\frac{1}{2}$}} \nc
{\arrow} [2] {\mbox{$\mathop{\rightarrow}\limits_{#1 \rightarrow
#2}$}}

\begin{document}

\title{Theoretical study of the direct $\alpha+d$ $\rightarrow$ $^6$Li + $\gamma $ astrophysical capture process in a three-body model II. Reaction rates and primordial abundance}

\author {E. M. Tursunov}
\email{tursune@inp.uz}
\affiliation{Curtin Institute for Computation and Department of Physics and Astronomy, Curtin University, GPO Box U1987, Perth, WA 6845, Australia}
\affiliation {Institute of Nuclear Physics, Academy of Sciences, 100214, Ulugbek, Tashkent, Uzbekistan}
\author {S. A. Turakulov}
\email{turakulov@inp.uz} \affiliation {Institute of Nuclear Physics,
Academy of Sciences, 100214, Ulugbek, Tashkent, Uzbekistan}
\author {A. S. Kadyrov}
\email{a.kadyrov@curtin.edu.au}
\affiliation{Curtin Institute for Computation and Department of Physics and Astronomy, Curtin University, GPO Box U1987, Perth, WA 6845, Australia}
\author {I. Bray}
\email{i.bray@curtin.edu.au}
\affiliation{Curtin Institute for Computation and Department of Physics and Astronomy, Curtin University, GPO Box U1987, Perth, WA 6845, Australia}

\begin{abstract}
 The astrophysical S-factor and reaction rate of the direct capture process $\alpha+d$ $\rightarrow$ $^6$Li + $\gamma$,
 as well as the abundance of the $^6$Li element are estimated in a three-body model.
 The initial state is factorized into the deuteron bound state and the $\alpha+d$ scattering state. The final nucleus
 $^6$Li(1+)  is described as a three-body bound state $\alpha+n+p$ in the hyperspherical Lagrange-mesh method.
Corrections to the asymptotics of the overlap integral in the S- and D-waves have been done for the E2 S-factor.
The isospin forbidden E1 S-factor is calculated from the initial isosinglet states to the small isotriplet
components of the final $^6$Li(1+) bound state. It is shown that the three-body model is able
to reproduce the newest experimental data of the LUNA collaboration for the astrophysical S-factor
and the reaction rates within the experimental error bars. The estimated $^6$Li/H abundance ratio of
$(0.67 \pm 0.01)\times 10^{-14}$ is in a very good agreement with the recent measurement $(0.80 \pm 0.18)\times 10^{-14}$ of the LUNA collaboration.
\end{abstract}


\maketitle

\section{Introduction}
\par There are two open astrophysical problems related to the abundance of lithium elements in the Universe. First, the 
Big Bang nucleosynthesis (BBN) model predicts for the  $^{7}$Li/H ratio an estimate about three times larger than the recent  astronomical observational data from metal-poor halo
stars \cite{sbor10,MSB16}.  The second lithium puzzle is related to the estimation of the primordial abundance ratio  $^{6}$Li/ $^{7}$Li
of the lithium isotopes. For this  ratio the BBN model \cite{serp04} yields a value about
three orders of magnitude less than the astrophysical data
\cite{asp06}. 
In the BBN model the abundance of the $^7$Li element is estimated on the basis of two key capture reactions    
$\alpha(^3$He,$\gamma)^7$Be and  $\alpha(^3$H,$\gamma)^7$Li  (see \cite{neff,navratil,tur18} and references therein).
For the estimation of the $^6$Li/$^7$Li ratio the BBN model includes
as input parameters the reaction rates of the direct radiative
capture process
\begin{eqnarray} \label{1}
\alpha+d\rightarrow {\rm ^6Li}+\gamma
\end{eqnarray}
at low energies within the range $30 \le E_{\rm cm} \le 400$ keV
\cite{serp04}. The data set of the LUNA collaboration at two
astrophysical energies E=94 keV and E=134 keV \cite{luna14} was
recently renewed with additional data at E=80 keV and E=120 keV  \cite{luna17}. These data sets were obtained as
results of the direct measurements of the astrophysical S-factor at
the underground facility. The new data are lower than the old data of
nondirect measurements from Ref. \cite{kien91}.
 Based on the new data set, the  thermonuclear reaction rate of the process has been estimated by the LUNA collaboration. 
 The results for the reaction rates turn out to be even lower than previously reported. 
 This further increases the discrepancy between prediction of the  BBN model
 and the astronomical observations for the  primordial abundance of the  $^6$Li element in the Universe \cite{luna17}.

 Until recently all the theoretical estimations of the astrophysical S-factor
 of the above direct capture reaction at low astrophysical energies
 were based on the so-called exact mass prescription, in the both potential models
 \cite{dub951,dub952,type97,desc98,mukh11,tur15,MSB16}
 and microscopic approaches \cite{lang86,noll01,TBL91}.
Within this prescription the matrix elements of the isospin
forbidden E1-transition were estimated by using the exact
experimental mass values of the colliding nuclei $^2$H and $^4$He.
As was shown recently in Ref. \cite{bt18} in details, this way has no
microscopic background at all and cannot be used, for example in
the description of the capture process $d(d,\gamma)^4$He of two
identical nuclei. Of course, the estimated in this way cross
sections and S-factors of the $\alpha(d,\gamma)^6$Li capture
reaction can be fortituously close to the experimental data, however
this method does not yield a relevant energy dependence of the
S-factor and cross section and correct predictive power for future
$\it ab-initio$ studies \cite{bt18}.
An alternative approach to the description of the capture processes is based on solving the three-body Faddeev equations 
 \cite{shub16} using quasi-separable potentials. An advantage of this method is that it allows an easier treatment of non-local effects that can be extended to three-body problems. 

Realistic three-body models are based on the isovector E1 transition
from the initial $T_i=0$  (isosinglet) states to the $T_f=1$
(isotriplet) components of the final $^6$Li$(1^+)$ bound state, or
from the initial isotriplet components to the main isoscalar part of
the final $^6$Li$(1^+)$ nucleus bound state \cite{bt18}. First
attempt to estimate in a correct way the matrix elements of the
isospin-forbidden E1- transition together with the E2-transition for
the $^4$He$(d,\gamma)^6$Li direct capture process has been done in
the three-body model \cite{TKT16}. The formalism of the model has
been developed in a consistent way and correct analytical
expressions have been obtained for the matrix elements of the E1-
and E2-transitions, including the isovector transition matrix
elements. The numerical results were obtained on the basis of the
final three-body wave function $^6$Li$=\alpha+p+n$ in hyperspherical
coordinates \cite{desc03,tur06}, which had a small isotriplet
component with the norm square of 1.13 $\times 10^{-5}$. 
Due to smallness of the isotriplet component of the final three-body bound state 
the corresponding numerical calculations in Ref. \cite{TKT16} have
 reproduced the existing experimental data for the S-factor only in the frame of the exact mass 
 prescription and with the help of additional spectroscopic factor.
Further studies in Ref. \cite{bt18} have demonstrated that the quality
of the final three-body wave function $^6$Li$=\alpha+p+n$ can be
improved and convergent isotriplet component can be reached with the
norm square of 5.3$\times 10^{-3}$, which is larger than the old
number by two orders of magnitude. This led to the fact that the E1
S-factor also increased by two orders of magnitude. Additionally, as
was shown in that paper, the E2 S-factor can be improved owing to
the correction of the asymptotics of the overlap integral of the
$^6$Li and deuteron wave functions at a distance 5-10 fm. 

The aim of present study is to 
estimate the reaction rates of the
$\alpha(d,\gamma)^6$Li direct capture process 
and the primordial
abundance of the $^6$Li element in the Universe
within the improved realistic three-body model \cite{TKT16,bt18}. The initial wave
function is factorized into the deuteron bound-state and the
$\alpha-d$ scattering-state wave functions. The final $^6$Li(1+)
state is described as a $\alpha+p+n$ three-body bound system. The
wave function on the hyperspherical Lagrange mesh basis available
for the $^6$Li(1+) bound state \cite{desc03,tur06} will be employed.

In Sec. II we describe the model, in Sec. III we discuss obtained
numerical results and finally, in the last section we draw
conclusions.

\section{Theoretical model}
\subsection{Cross sections of the radiation capture process}

The cross sections of the radiative capture process reads
\begin{align}
\sigma_{E}(\lambda)=& \sum_{J_i T_i \pi_i}\sum_{J_f T_f
\pi_f}\sum_{\Omega \lambda}\frac{(2J_f+1)} {\left [I_1
\right]\left[I_2\right]} \frac{32 \pi^2 (\lambda+1)}{\hbar \lambda
\left( \left[ \lambda \right]!! \right)^2} k_{\gamma}^{2 \lambda+1} C_s^2
\nonumber \\ &\times \sum_{l_\omega I_\omega}
 \frac{1}{ k_\omega^2 v_\omega}\mid
 %<
 \langle \Psi^{J_f T_f \pi_f}
% \nonumber \\ &\times
\|M_\lambda^\Omega\|
\Psi_{l_\omega I_\omega}^{J_i T_i \pi_i}
%>
\rangle \mid^2,
\end{align}
where $\Omega=$E  or M (electric or magnetic transition), $\omega$
denotes the entrance channel, $k_{\omega}$, $v_\omega$,  $I_\omega$
are the wave number, velocity of the $\alpha-d$ relative motion and
the spin of the entrance channel, respectively, $J_f$, $T_f$,
$\pi_f$ ($J_i$, $T_i$, $\pi_i$) are the spin, isospin and parity of
the final (initial) state, $I_1$, $I_2$ are channel spins,
$k_{\gamma}=E_\gamma / \hbar c$ is the wave number of the photon
corresponding to the energy $E_\gamma=E_{\rm th}+E$ with the
threshold energy $E_{\rm th}=1.474$ MeV. The wave functions
$\Psi_{l_\omega I_\omega}^{J_i T_i \pi_i} $ and $\Psi^{J_f T_f
\pi_f} $ represent  the initial and final states, respectively. The
reduced matrix elements are evaluated between the initial and final
states. We also use short-hand notations $[I]=2I+1$ and
$[\lambda]!!=(2\lambda+1)!!$. 

Constant $C_s^2$ is the spectroscopic factor
\cite{Angulo}. As argued in Ref. \cite{mukh11}, if the two-body potentials of the model correctly reproduce experimental phase shifts in the partial waves
and physical bound state energies of the two-body subsystems, then     
a value of the spectroscopic factor must be taken equal to 1.  This reflects the fact that the potential parameters already include many-body effects. Accordingly, the factor is set equal to 1.

The analytical expressions of the E1 and E2 electric-transition
operators, including isospin transition operators and their matrix
elements in the three-body model have been described in Ref.
\cite{TKT16}. For the sake of brevity we refer to that paper for the
details of the model.

The astrophysical $S$-factor of the process is expressed in terms of  the
cross section as \cite{Fowler}
\begin{eqnarray}
S(E)=E \, \sigma_E(\lambda) \exp(2 \pi \eta),
\end{eqnarray}
where  $\eta$ is the Coulomb parameter.

\subsection{Reaction rates}

The reaction rate $N_{a}(\sigma v)$ is estimated according to
\cite{Fowler, Angulo}
\begin{align}
N_{a}(\sigma v)=N_{A}
\frac{(8/\pi)^{1/2}}{\mu^{1/2}(k_{\text{B}}T)^{3/2}}
\int^{\infty}_{0} \sigma(E) E \exp(-E/k_{\text{B}}T) d E,
\end{align}
where $k_{\text{B}}$ is the Boltzmann constant, $T$ is the
temperature, $N_{A}=6.0221\times10^{23}\, \text{mol}^{-1}$ is the
Avogadro number. The reduced mass is written as $\mu=A m_N$ with the corresponding reduced mass
number $A=A_1 A_2/(A_1 + A_2)$ for the $\alpha+d$ system, where
$A_1=2$ and $A_2=4$. Consequently, a value of $A=4/3$ is fixed. When
a variable $k_{\text{B}}T$ is expressed in units of MeV it is
convenient to use a variable $T_9$ for the temperature in units of
$10^9$ K according to the equation $k_{\text{B}}T=T_{9}/11.605$ MeV.
In our calculations $T_9$ varies in the interval $0.001\leq T_{9}
\leq 10$.

After substitution of these variables the above integral for the
reaction rates can be expressed as:
\begin{eqnarray}
\label{rate}
 N_{a}(\sigma v)&=&3.7313 \times 10^{10}A^{-1/2}\,\, T_{9}^{-3/2}
 \nonumber \\ &&\times
\int^{ \infty}_{0} \sigma(E) E \exp(-11.605E/T_{9}) d E.
\end{eqnarray}

\section{Numerical results}
\subsection{Details of the calculations}

Calculations of the cross section and astrophysical S-factor have
been performed under the same conditions as in Ref.\cite{bt18}. The
radial wave function of the deuteron is the solution of the
bound-state Schr{\"o}dinger equation with the central Minnesota
potential $V_{NN}$ \cite{thom77,RT70} with $\hbar^2/2
m_N=20.7343$ MeV fm$^2$. The Schr{\"o}dinger equation is solved
using a highly accurate Lagrange-Laguerre mesh method \cite{baye15}.
It yields $E_d$=-2.202 MeV for the deuteron ground-state energy with
the number of mesh points $N=40$ and a scaling parameter $h_d=0.40$.

The scattering wave function of the $\alpha-d$ relative motion is
calculated with a deep potential of Dubovichenko \cite{dub94} with a
small modification in the $S$-wave \cite{tur15}:
$V_d^{(S)}(R)=-92.44 \exp(- 0.25 R^2) $ MeV. The potential
parameters in the $^3P_0$, $^3P_1$, $^3P_2$  and $^3D_1$, $^3D_2$,
$^3D_3$ partial waves are the same as in Ref. \cite{dub94}. The
potential contains additional states in the $S$- and $P$-waves
forbidden by the Pauli principle. The above modification of the
S-wave potential reproduces the empirical value $C_{\alpha d}=2.31$
fm$^{-1/2}$ of the asymptotic normalization coefficient (ANC) of the
$^6$Li(1+) ground state derived from $\alpha-d$ elastic scattering
data \cite{blok93}.

The final $^6$Li(1+) ground-state wave function was calculated using
the hyperspherical Lagrange-mesh method \cite{desc03,tur06,tur07}
with the same Minnesota NN-potential. For the $\alpha-N$ nuclear
interaction the potentials of Voronchev {\em et al.}  (Model A)
\cite{vor95} and Kanada {\em et al.} (Model B) \cite{KKN79} were
employed, which contain a deep Pauli-forbidden state in the
$S$-wave. The potentials were slightly renormalized by a scaling
factors 1.014 (Model A) and 1.008 (Model B) to reproduce the
experimental binding energy $E_b$=3.70 MeV. The Coulomb interaction
between $\alpha$ and proton is taken as $2e^2\, \mathrm{erf}(0.83\,
R)/R$ \cite{RT70}. The coupled hyperradial equations are solved with
the Lagrange-mesh method \cite{baye15,desc03}. The hypermomentum
expansion includes terms up to $K_{\rm max} = 24$, which ensures a
good convergence of the energy and of the $T = 1$ component of
$^6$Li. The ground state is essentially $S = 1$ (96 \%). The matter
r.m.s. radius of the ground state (with 1.4 fm as $\alpha$ radius)
is found as $\sqrt{r^2} \approx 2.25$ fm with the potential of
\re{vor95} or 2.24 fm with the potential of \re{KKN79},
\textit{i.e.}\ values slightly lower than the experimental value
$2.32 \pm 0.03$ fm \cite{exp162}. The isotriplet component in the
$^6$Li ground state has a squared norm $5.3 \times 10^{-3}$ with the
potential of \re{vor95} (Model A) and $4.2 \times 10^{-3}$ with the
potential of \re{KKN79} (Model B).

\subsection{Estimation of the astrophysical S-factor}

\begin{figure}[htb]
%\centering
\includegraphics[width=\columnwidth]{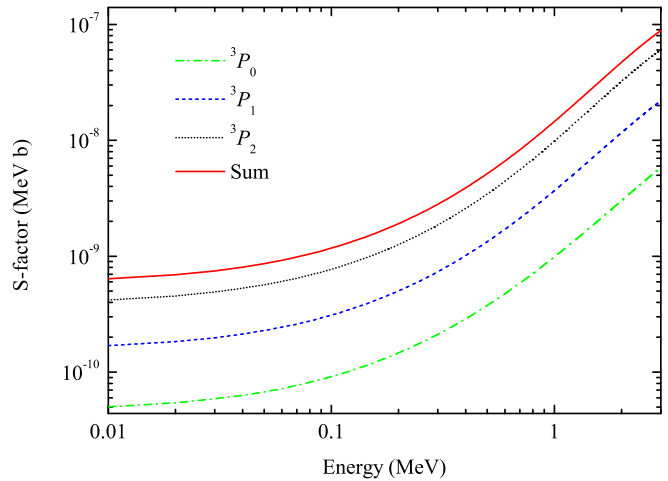}
\caption{Partial E1 astrophysical S-factors of the direct
$\alpha+d\rightarrow ^6$Li$+\gamma$ capture process within Model
A (see text).} \label{FIG1}
\end{figure}

In Fig. \ref{FIG1} we display E1 astrophysical S-factors for the
direct $\alpha+d\rightarrow ^6$Li$+\gamma$ capture process within
Model A from the initial partial  $^3P_0$, $^3P_1$, $^3P_2$
scattering waves to the $T$=1 (isotriplet) components of the final
ground state of $^6$Li.

At low astrophysical energies, the cross section is very sensitive
to the asymptotic behavior of the overlap integrals of the deuteron wave function
$\Psi_d$ and the three-body wave functions $\Psi_f^{1M+}$  for
 the $L = 0$ and $L=2$ partial waves up to large $\alpha-d$ distances
$R$.  The asymptotic normalization coefficients (ANCs) of the $^6$Li nucleus in the $\alpha+d$ channel can be extracted 
within the effective range expansion method \cite{blokh17,blokh18} or  from the analytical continuation of the scattering amplitude \cite{blok93}.

The overlap integrals are written as 
 \beq I_L(R) = \langle [\Psi_d \otimes Y_L(\Omega_R)]^{1M} |
\Psi_f^{1M+} \rangle, \eeqn{3.50}  
where the integration is done over internal coordinates of the
deuteron and the angular part of the variable~$\bf{R}$. 
In the
present three-body model, over the interval
$5-10$ fm $I_L(R)$ follows  the expected asymptotic behavior $C_{\alpha d}^{(L)}
W_{-\eta_b,L+1/2}(2k_bR)/R$, where $\eta_b$ and $k_b$ are the
Sommerfeld parameter and wave number calculated at the separation
energy 1.474 MeV of the $^6$Li bound state into $\alpha$ and $d$
\cite{bt18}. The values of the S-wave and D-wave asymptotic
normalization coefficients (ANC) have been estimated for different
values of matching point $R_0$. We found that S-wave  ANC is maximal
(consequently optimal) for the matching point at 5.5 fm: $ C_{\alpha
d}^{(0)}=2.116$ fm$^{-1/2}$ and $C_{\alpha d}^{(0)}=2.051$
fm$^{-1/2}$ for Model A and Model B, respectively. The first
number is slightly larger than $C_{\alpha d}^{(0)} \approx 2.05$
fm$^{-1/2}$ \cite{bt18}, obtained with $R_0=7.75$ fm and in
reasonable agreement with the value $C_{\alpha d}^{(0)} \approx
2.30$ fm$^{-1/2}$ extracted in \re{blok93} from experimental data on
$\alpha+d$ scattering. The estimated values of D-wave ANC  are less
than the corresponding values of the S-wave ANC by two orders of
magnitude and vary in the range between $2.160\times 10^{-2}$ and $2.175\times 10^{-2}$ 
fm$^{-1/2}$ for model A for matching points from $R_0=$5.5 fm to
7.5 fm. Model B yields the range between $2.179\times 10^{-2}$ and $2.188\times 10^{-2}$
 fm$^{-1/2}$, respectively.

\begin{figure}[htb]
\includegraphics[width=\columnwidth]{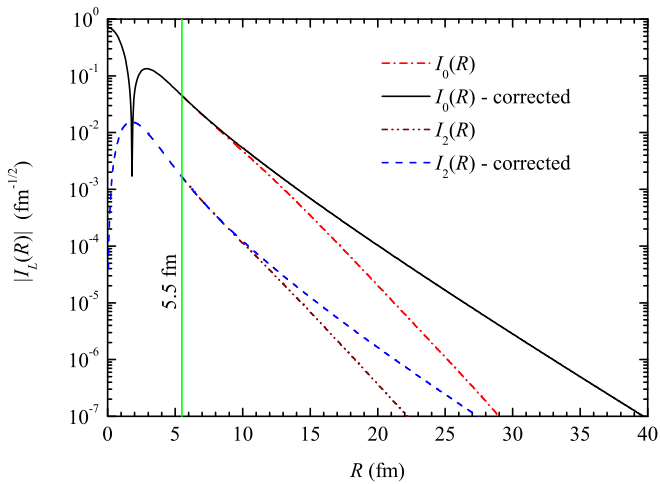}
\caption{Overlap integral with the initial three-body and the corrected
(at $R_0$=5.5 fm) asymptotics within Model A.} \label{FIG2}
\end{figure}

In Fig. \ref{FIG2} the overlap integrals $I_0(r)$ and $I_2(r)$ with
the initial three-body and the asymptotics corrected at $R_0=5.5$ fm, 
within Model A are displayed. The S-wave overlap integral changes
the sign at small distances due to orthogonality to the $\alpha - d$
Pauli-forbidden state, this is why the absolute values of the overlap integrals are shown. 
As can be seen from the figure, beyond about 10 fm the
absolute value of $I_L(R)$ decreases faster than the correct
asymptotics. Hence, within the three-body model, the E2
astrophysical S-factor is underestimated at low collision energies.
This is the motivation to estimate the E2 S-factor with corrected
asymptotics of the overlap integral.

\begin{figure}[htb]
\includegraphics[width=\columnwidth]{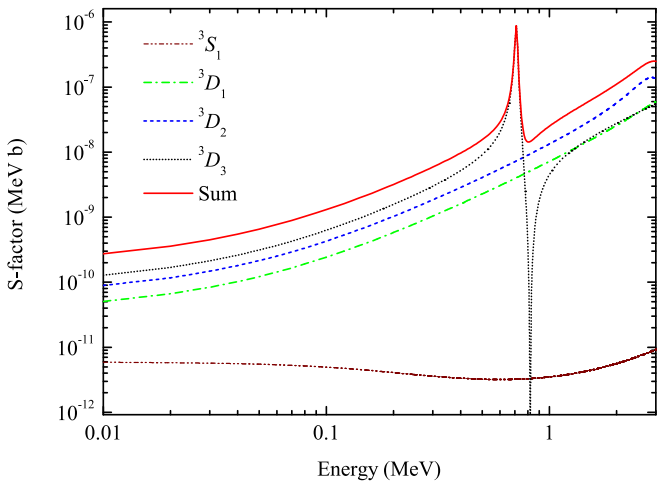}
\caption{Partial E2 astrophysical S-factors of the direct
$\alpha+d\rightarrow ^6$Li$+\gamma$ capture process within Model
A with the corrected asymptotics of the overlap integral.} \label{FIG3}
\end{figure}

In Fig. \ref{FIG3} we show E2 astrophysical S-factors for the direct
$\alpha+d\rightarrow ^6$Li$+\gamma$ capture process within Model
A from the initial $^3S_1$, $^3D_1$, $^3D_2$, $^3D_3$ partial waves
to the ground state of $^6$Li with the corrected asymptotics of
$I_0(R)$ and $I_2(R)$ at a distance $R_0 = 5.5$ fm. As can be  seen
from the figure, at low energies the contribution of the partial $^3S_1$ $\alpha+d$
configuration is less than the contributions of partial D-waves at
least by an order of magnitude. However, the S-wave
contribution has a weak energy dependence,
%decreasing slightly within the interval from 
%$3.0\times 10^{-12}$  to $6.0\times 10^{-12}$ MeV b, 
while the
smallest $^3D_1$ wave contribution increases sharply from $5\times 10^{-11}$ up
to $6\times 10^{-8}$ MeV b within the same energy interval.

\begin{figure}[htb]
\includegraphics[width=\columnwidth]{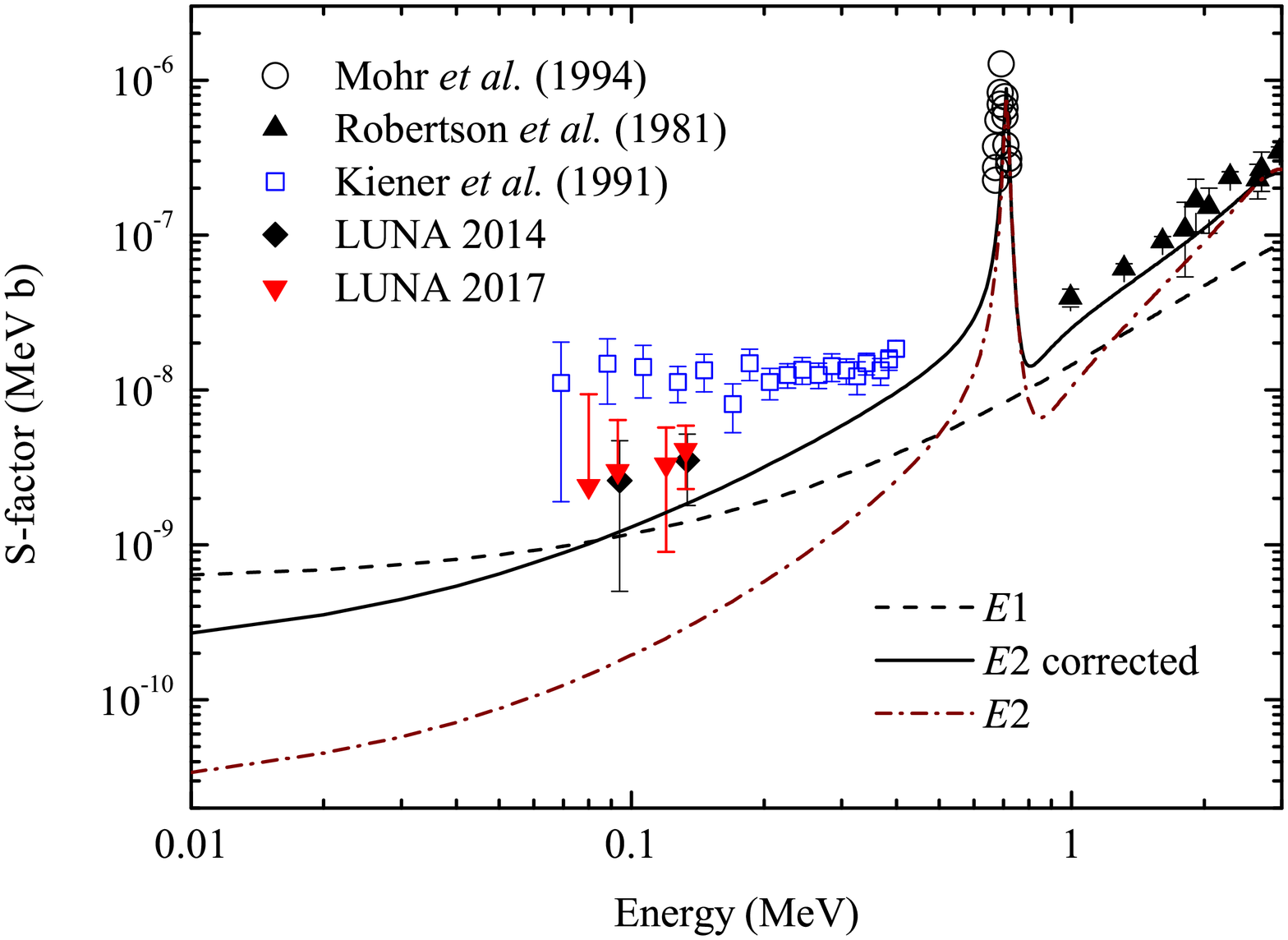}
\caption{Relative contributions of the E1 and E2 astrophysical
S-factors of the direct $\alpha+d\rightarrow ^6$Li$+\gamma$ capture
process within Model A in comparison with available experimental
data.} \label{FIG4}
\end{figure}

\begin{figure}[htb]
\includegraphics[width=\columnwidth]{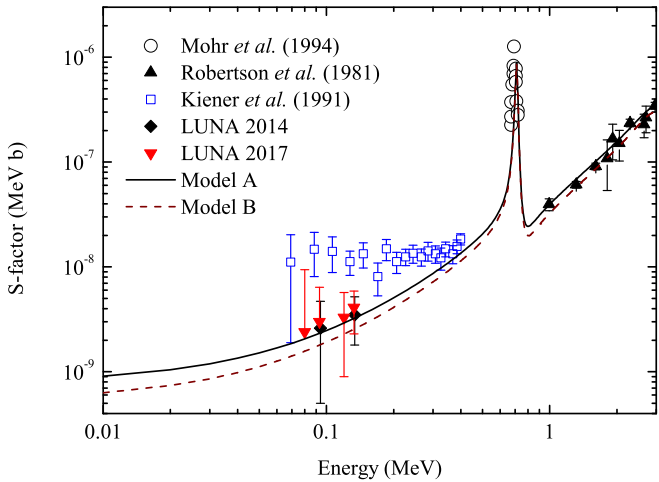}
\caption{Theoretical astrophysical S-factor for the direct
$\alpha+d\rightarrow ^6$Li$+\gamma$ capture process within
Models A and B in comparison with available experimental data.}
\label{FIG5}
\end{figure}

In Fig.~\ref{FIG4}  we compare the E1 and E2 transition components
of the S-factor with available experimental data, including recent
data from Refs. \cite{luna14,luna17}. As can be seen from the
figure, at low energies the E1 transition dominates even with
corrected asymptotics of the overlap integral for the E2 transition,
while at higher energies the E2 component is stronger.
Finally, in Fig.~\ref{FIG5} we compare the obtained theoretical
results for the astrophysical S-factor of the direct
$\alpha+d\rightarrow ^6$Li$+\gamma$ capture process with
experimental data from Refs.~\cite{mohr94,robe81,kien91,luna14,luna17}.
One can note that Figs.~\ref{FIG4} and \ref{FIG5} are very similar to
Figs.~\ref{FIG1} and \ref{FIG2} of Ref.~\cite{bt18}, respectively. In
fact, presently we include also a correction to the D-wave
asymptotics of the overlap integral. Indeed, due to small values of
the D-wave ANC of order 10$^{-2}$, the corresponding E2 S-factor is
very small and one can not see its difference from the results of
Ref. \cite{bt18}. However, even small D-wave corrections can give a
non-negligible contribution to the reaction rates of the process.

As was noted in Ref.~\cite{bt18}, the E2 S-factor can be enhanced
owing to the D-wave components of the deuteron, $^4$He and the final
$^6$Li nucleus with the help of tensor forces in microscopic $\it
ab-initio$ models. Together with the aforementioned weak dependence of the E2 S-factor
from the initial S-wave at very low energies this can lead to a
larger S-wave contribution for the process at low astrophysical
energies.

\subsection{Reaction rates and abundance}

In Table \ref{tab01} we give theoretical estimations for the
$d(\alpha,\gamma)^{6}$Li reaction rates in the temperature interval
 $10^{6}$ K $\leq T \leq
10^{10}$ K ($ 0.001\leq T_{9} \leq 10 $) calculated with the two
$\alpha+N$ potentials of Voronchev {\em et al.}  \cite{vor95} (Model A) and
Kanada et al \cite{KKN79} (Model B). In the second and third columns
of the table we give "the most effective energy" $E_0$  and the
width of the Gamov window $\Delta E_0$ (\ref{rate}). They
are expressed as \cite{Angulo}:
\begin{eqnarray}
E_0&=&\left( \frac{\mu}{2} \right)^{1/3} \left( \frac{\pi e^2 Z_1Z_2
k_B T}{\hbar} \right)^{2/3} 
\nonumber \\
&=& 0.122 \, (Z_1^2 Z_2^2 A)^{1/3} T_9
^{2/3} \, [\textrm {MeV}],
\end{eqnarray}
and
\begin{eqnarray}
\Delta E_0&=&4 \left( E_0 k_B T/3 \right)^{1/2}
\nonumber \\
&=&  0.2368 \,
(Z_1^2 Z_2^2 A)^{1/6} T_9 ^{5/6} \, [\textrm {MeV}].
\end{eqnarray}

%\clearpage
\begin{widetext}

\begin{table}[htb] 
\caption{Theoretical estimations of the direct ${d}(\alpha, \gamma)^{6}{\rm Li}$ capture
reaction rate in the temperature interval
$
%\,\,\,\,\,\,\,\,\,\,\,\,\,\,\,\,\,\,\,\,\,\,\, 
10^{6}$ K $\leq T
\leq 10^{10}$ K ($ 0.001\leq T_{9} \leq 10 $).}
\label{tab01}
\begin{center}
{%\scriptsize

%\begin{tabular}{|c|c|c|c|c|c|c|c|c|c|}
\begin{tabular}{ccccccccccc}
\hline \hline
$T_9$ & $E_0$ (MeV) & $\Delta E_0$ (MeV) &
\multicolumn{2}{c}{$N_{a}(\sigma v)$ ($ \textrm{cm}^{3}
\rm{mol}^{-1} \rm{s}^{-1}$)} &$\,\,\,\,\,\,\,$& $T_9$ & $E_0$ (MeV) & $ \Delta E_0$ (MeV)
& \multicolumn{2}{c}{$N_{a}(\sigma v)$ ($ \textrm{cm}^{3}
\rm{mol}^{-1} \rm{s}^{-1}$)} \\ %\hline
 &  &  & \textrm{Model A} & \textrm{Model B} &  &&  &  & \textrm{Model A} & \textrm{Model B}\\
 \hline
0.001 & 0.002 & 0.001 & $3.47\times10^{-30}$ & $2.37\times10^{-30}$ && 0.120 & 0.052 & 0.054 & $1.83\times10^{-5}$ & $1.38\times10^{-5}$\\
0.002 & 0.003 & 0.002 & $1.04\times10^{-23}$ & $7.14\times10^{-24}$ && 0.130 & 0.055 & 0.057 & $2.68\times10^{-5}$ & $2.03\times10^{-5}$ \\
0.003 & 0.004 & 0.003 & $1.42\times10^{-20}$ & $9.74\times10^{-21}$ && 0.140 & 0.058 & 0.061 & $3.79\times10^{-5}$ & $2.88\times10^{-5}$\\
0.004 & 0.005 & 0.003 & $1.33\times10^{-18}$ & $9.18\times10^{-19}$ && 0.150 & 0.060 & 0.064 & $5.21\times10^{-5}$ & $3.96\times10^{-5}$  \\
0.005 & 0.006 & 0.004 & $3.36\times10^{-17}$ & $2.32\times10^{-17}$ && 0.160 & 0.063 & 0.068 & $6.96\times10^{-5}$ & $5.31\times10^{-5}$ \\
0.006 & 0.007 & 0.004 & $3.93\times10^{-16}$ & $2.72\times10^{-16}$ && 0.180 & 0.068 & 0.075 & $1.17\times10^{-4}$ & $8.96\times10^{-5}$ \\
0.007 & 0.008 & 0.005 & $2.79\times10^{-15}$ & $1.93\times10^{-15}$ && 0.200 & 0.073 & 0.082 & $1.83\times10^{-4}$ & $1.41\times10^{-4}$ \\
0.008 & 0.009 & 0.006 & $1.41\times10^{-14}$ & $9.77\times10^{-15}$ && 0.250 & 0.085 & 0.099 & $4.53\times10^{-4}$ & $3.54\times10^{-4}$ \\
0.009 & 0.009 & 0.006 & $5.52\times10^{-14}$ & $3.83\times10^{-14}$ && 0.300 & 0.096 & 0.115 & $9.17\times10^{-4}$ & $7.23\times10^{-4}$ \\
0.010 & 0.010 & 0.007 & $1.79\times10^{-13}$ & $1.25\times10^{-13}$ && 0.350 & 0.106 & 0.131 & $1.62\times10^{-3}$ & $1.29\times10^{-3}$ \\
0.011 & 0.011 & 0.007 & $5.00\times10^{-13}$ & $3.48\times10^{-13}$ && 0.400 & 0.116 & 0.146 & $2.62\times10^{-3}$ & $2.10\times10^{-3}$ \\
0.012 & 0.011 & 0.008 & $1.24\times10^{-12}$ & $8.66\times10^{-13}$ && 0.500 & 0.134 & 0.176 & $5.68\times10^{-3}$ & $4.60\times10^{-3}$ \\
0.013 & 0.012 & 0.008 & $2.80\times10^{-12}$ & $1.96\times10^{-12}$ && 0.600 & 0.152 & 0.205 & $1.06\times10^{-2}$ & $8.67\times10^{-3}$ \\
0.014 & 0.012 & 0.009 & $5.82\times10^{-12}$ & $4.08\times10^{-12}$ && 0.700 & 0.168 & 0.233 & $1.79\times10^{-2}$ & $1.49\times10^{-2}$ \\
0.015 & 0.013 & 0.010 & $1.13\times10^{-11}$ & $7.94\times10^{-12}$ && 0.800 & 0.184 & 0.260 & $2.88\times10^{-2}$ & $2.43\times10^{-2}$ \\
0.016 & 0.014 & 0.010 & $2.08\times10^{-11}$ & $1.46\times10^{-11}$ && 0.900 & 0.199 & 0.287 & $4.43\times10^{-2}$ & $3.80\times10^{-2}$ \\
0.018 & 0.015 & 0.011 & $6.11\times10^{-11}$ & $4.30\times10^{-11}$ && 1.000   & 0.213 & 0.313 & $6.56\times10^{-2}$ & $5.70\times10^{-2}$ \\
0.020 & 0.016 & 0.012 & $1.55\times10^{-10}$ & $1.09\times10^{-10}$  && 1.500 & 0.279 & 0.439 & $2.72\times10^{-1}$ & $2.45\times10^{-1}$\\
0.025 & 0.018 & 0.015 & $9.90\times10^{-10}$ & $7.03\times10^{-10}$ && 2.000   & 0.338 & 0.558 & $6.04\times10^{-1}$ & $5.50\times10^{-1}$\\
0.030 & 0.021 & 0.017 & $4.08\times10^{-9}$ & $2.91\times10^{-9}$    && 2.500 & 0.393 & 0.672 & $9.88\times10^{-1}$ & $8.99\times10^{-1}$ \\
0.040 & 0.025 & 0.021 & $3.23\times10^{-8}$ & $2.32\times10^{-8}$    && 3.000 & 0.443 & 0.782 & $1.39$ & $1.26$\\
0.050 & 0.029 & 0.026 & $1.41\times10^{-7}$ & $1.02\times10^{-7}$    && 4.000 & 0.537 & 0.994 & $2.26$ & $2.02$\\
0.060 & 0.033 & 0.030 & $4.35\times10^{-7}$ & $3.18\times10^{-7}$    && 5.000 & 0.623 & 1.197 & $3.24$ & $2.87$\\
0.070 & 0.036 & 0.034 & $1.07\times10^{-6}$ & $7.88\times10^{-7}$    && 6.000 & 0.704 & 1.393 & $4.35$ & $3.83$\\
0.080 & 0.040 & 0.038 & $2.26\times10^{-6}$ & $1.67\times10^{-6}$    && 7.000 & 0.780 & 1.584 & $5.54$ & $4.87$\\
0.090 & 0.043 & 0.042 & $4.27\times10^{-6}$ & $3.17\times10^{-6}$    && 8.000 & 0.853 & 1.771 & $6.78$ & $5.95$\\
0.100 & 0.046 & 0.046 & $7.38\times10^{-6}$ & $5.51\times10^{-6}$    && 9.000 & 0.922 & 1.953 & $8.05$ & $7.06$\\
0.110 & 0.049 & 0.050 & $1.19\times10^{-5}$ & $8.94\times10^{-6}$    && 10.00 & 0.989 & 2.133 & $9.31$ & $8.16$ \\
\hline \hline
\end{tabular}
}
\end{center}
\end{table}
\begin{table}[htb]
\caption{The fitting coefficients of the analytical
approximation for the direct  ${d}(\alpha,
\gamma)^{6}{\rm Li}$ capture reaction rate.}
\label{tab02}
\begin{center}
{%\scriptsize

%\begin{tabular}{|c|c|c|c|c|c|c|c|c|c|}
\begin{tabular}{cccccccccc}
\hline \hline
\textrm{Model} & $p_0$ & $p_1$ & $p_2$ & $p_3$ & $p_4$ & $p_5$ & $p_6$ & $p_7$ & $p_8$ \\
\hline
\textrm{ A} &6.004 &-2.558 &34.730 &-115.482& 205.801&-169.456 & 71.428 & -11.614&42.354 \\
\textrm{ B} &5.154 &-5.830 &52.356 &-163.500& 272.839&-218.444 & 89.174 & -14.107&41.384 \\

\hline \hline
\end{tabular}
}
\end{center}
\end{table}
\end{widetext}

\begin{figure}[htb]
\includegraphics[width=\columnwidth]{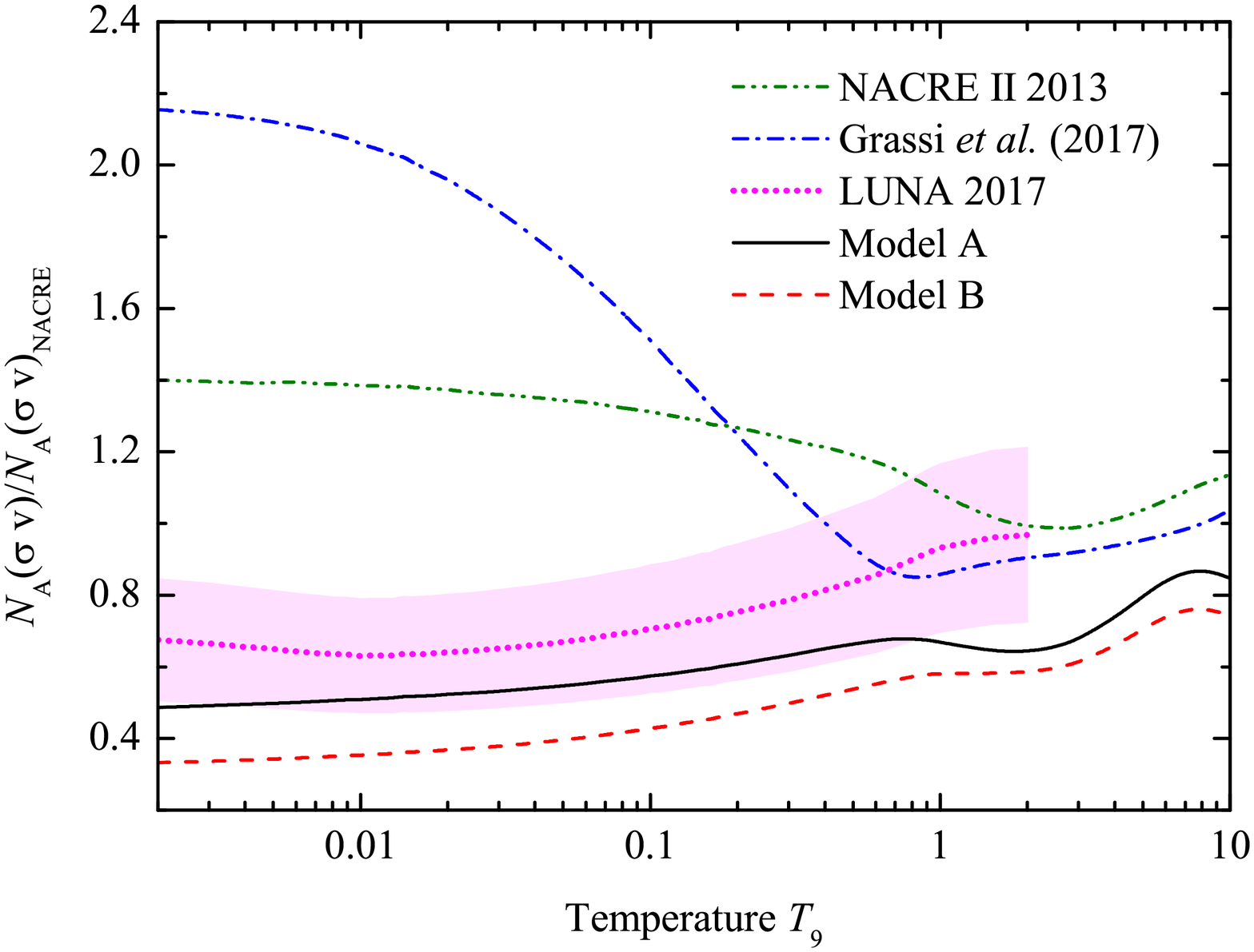}
\caption{Reaction rates of the direct $\alpha+d\rightarrow
^6$Li$+\gamma$ capture process within Models A and B normalized
to the NACRE 1999 experimental data.} \label{FIG6}
\end{figure}

In Fig. \ref{FIG6} we display the estimated reaction rates of the
direct $\alpha+d\rightarrow ^6$Li$+\gamma$ capture process within
Models A and B normalized to the standard NACRE 1999
experimental data \cite{NACRE99}. For comparison we also display
the lines corresponding to the adopted values of the NACRE II 2013 data \cite{NACRE13},
new LUNA 2017 \cite{LUNA17} data and data fit from
Ref. \cite{Laura17}. As can be seen from the figure, our results
obtained within Models A and B show the same temperature
dependence at low values of $T_9$, as the newest direct data of the
LUNA 2017 \cite{LUNA17} and differs from the data NACRE II
2013 \cite{NACRE13} and the data fit in Ref. \cite{Laura17}.
Consequently, the corresponding energy dependence of the astrophysical S-factor,
obtained in the developed theoretical model is mostly consistent with the last direct data
of the LUNA collaboration \cite{LUNA17}.

For the estimation of the abundance of the $^6$Li element, the theoretical reaction rate is approximated within 1.84\% (Model A) and 2.46  \% (Model B) using the following analytical formula:
\begin{eqnarray}
 N_{24}(\sigma v)&=&p_0 T_{9}^{-2/3} \exp ( -7.423 T_{9}^{-1/3}) \nonumber \\
 && \times \left[ 1 + p_1 T_9^{1/3} + p_2 T_{9}^{2/3}+p_3 T_{9}+p_4 T_{9}^{4/3} \right. \nonumber \\
 && 
+ \left. p_5 T_{9}^{5/3}+p_6 T_{9}^2+ p_7 T_{9}^{7/3} \right] \nonumber \\ 
&&+ p_8 T_{9}^{-3/2} \exp (-7.889 T_9^{-1}).
\end{eqnarray}
The coefficients of the analytical polynomial
approximation of the $d(\alpha,\gamma)^{6}$Li reaction rates
estimated with the $\alpha+N$ potential of Voronchev {\em et al.} (Model A) and  Kanada {\em et al.} (Model B)  are given in Table \ref{tab02} in the temperature interval ($0.001\leq T_{9} \leq 10 $).

On the basis of the theoretical reaction rates and with the help of the
PArthENoPE \cite{Pisanti08} public code we have estimated the
primordial abundance of the $^6$Li element. If we adopt the Planck
2015 best fit for the baryon density parameter
  $\Omega_b h^2=0.02229^{+0.00029}_{-0.00027}$  \cite{ade16}   and the neutron life time $\tau_n=880.3 \pm
1.1$ s  \cite{olive14}, for the $^6$Li/H abundance ratio we have an estimation from $0.66\times 10^{-14}$ to $0.68\times 10^{-14}$ within Model A.
Model B yields an estimation from $0.49\times 10^{-14}$ to $0.51\times 10^{-14}$. The results of Model A are mostly consistent with the new estimation $^6$Li/H=$(0.80 \pm 0.18)\times 10^{-14}$ of the LUNA collaboration \cite{LUNA17} than the models
based on the exact mass prescription method \cite{Laura17} $^6$Li/H=$(0.90 - 1.8)\times 10^{-14}$.  
Finally, using this result and the estimate of the $^7$Li/H abundance ratio of  $(5.2 \pm 0.4)\times 10^{-10}$ from Ref. \cite{kontos13} we get $^6$Li/$^7$Li=$(1.30 \pm 0.12)\times 10^{-5}$ which agrees with the standard estimate from the BBN model \cite{serp04}.

\section{Conclusions}

The astrophysical direct capture process  $\alpha+d\rightarrow
^6$Li$+\gamma$ has been studied in the three-body model. The reaction rates, E1 and
E2 astrophysical S-factors as well as  
the primordial abundance of the $^6$Li element have been
estimated. 
The asymptotics of the overlap integral in
the S- and D-waves have been corrected. This increased the E2 S-factor by an order of magnitude at low astrophysical energies. 
Together  with the corrected 
E2 S-factor, the contribution of the E1-transition operator to the
S-factor from the initial isosinglet states to the small isotriplet
components of the final $^6$Li(1+) bound state is shown to be able
to reproduce the new experimental data of the LUNA collaboration
within the experimental error bars. The theoretical reaction rates
have the same temperature dependence at low temperatures as the
newest direct 2017 data of the LUNA collaboration. For the abundance
ratio $^6$Li/H we have obtained an estimation $(0.67 \pm 0.01)\times 10^{-14}$ , 
consistent with the new estimation of the LUNA collaboration  and much lower than the 
results of the models based on the exact mass prescription. Further
improvement of the theoretical estimations of the reaction rates and
$^6$Li abundance is expected with the help of NN-tensor forces within
{\it ab-initio} calculations.

%\section*{Acknowledgements}
\acknowledgments

The authors acknowledge Daniel Baye and Pierre Descouvemont for useful discussions and valuable advice. 
E.M.T. acknowledges a visiting scholarship from the Curtin Institute for Computation
and thanks members of the Theoretical physics group at
Curtin University for the kind hospitality during his visit. 
The support of the Australian Research Council, the
Australian National Computer Infrastructure, and the Pawsey
Supercomputer Centre are gratefully acknowledged.

%\bibliography{myreference}

\end{document}